\documentclass[usenatbib]{iaus}
\usepackage{graphicx}

\title[The Effect of Binary Interactions in IR Passbands]
{The Effect of Binary Interactions in Infrared Passbands}

\author[F. Zhang, L. Li \& Z. Han]
{F. Zhang$^1$, L. Li$^1$ \and Z. Han$^1$}

\affiliation{$^1$National Astronomical Observatories/Yunnan
Observatory, Chinese Academy of Sciences, PO Box 110, Kunming,
Yunnan Province 650011, China \break email:
gssephd@public.km.yn.cn .or. zhangfh@ynao.ac.cn}

\pubyear{2008}
\volume{252}
\pagerange{119--126}
\date{?? and in revised form ??}
\jname{The Art of Modelling Stars in the 21st Century}
\editors{Licai Deng, K. L. Chan \& C. Chiosi, eds.}
\begin{document}

\maketitle

\begin{abstract}
We present the integrated $J, H, K, L, M$ and $N$ magnitudes and
the colours involving infrared bands, for an extensive set of
instantaneous-burst binary stellar populations (BSPs) by using
evolutionary population synthesis (EPS). By comparing the results
for BSPs {\it WITH} and {\it WITHOUT} binary interactions we show
that the inclusion of binary interactions makes the magnitudes of
populations larger (fainter) and the integrated colours smaller
(bluer) for $\tau \ge 1$\,Gyr.
Also, we compare our model magnitudes and colours with those of
Bruzual \& Charlot (2003, hereafter BC03) and Maraston (2005,
hereafter M05). At last, we compare these model broad colours with
Magellanic Clouds globular clusters (GCs) and Milky Way GCs. In
$(V-R)-$[Fe/H] and $(V-I)-$[Fe/H] diagrams it seems that our
models match the observations better than those of BC03 and M05.

\end{abstract}

{\bf Introduction} In previous paper (Zhang et al. 2005) we took
into account binary interactions (BIs) in evolutionary population
synthesis (EPS) models, presented the integrated $U-B$, $B-V$,
$V-R$ and $V-I$ colours of binary stellar populations (BSPs),
while did not give the infrared magnitudes and colours because
larger fluctuations exist.
However, these results in infrared passbands are very important in
EPS models because the infrared light can reflect the metallicity
of populations and the visible/infrared colours are the candidates
of breaking the degeneration between age and metallicity.

\begin{figure}
\centering{
\includegraphics[bb=79 37 581 702,height=3.0cm,width=2.6cm,clip,angle=-90]{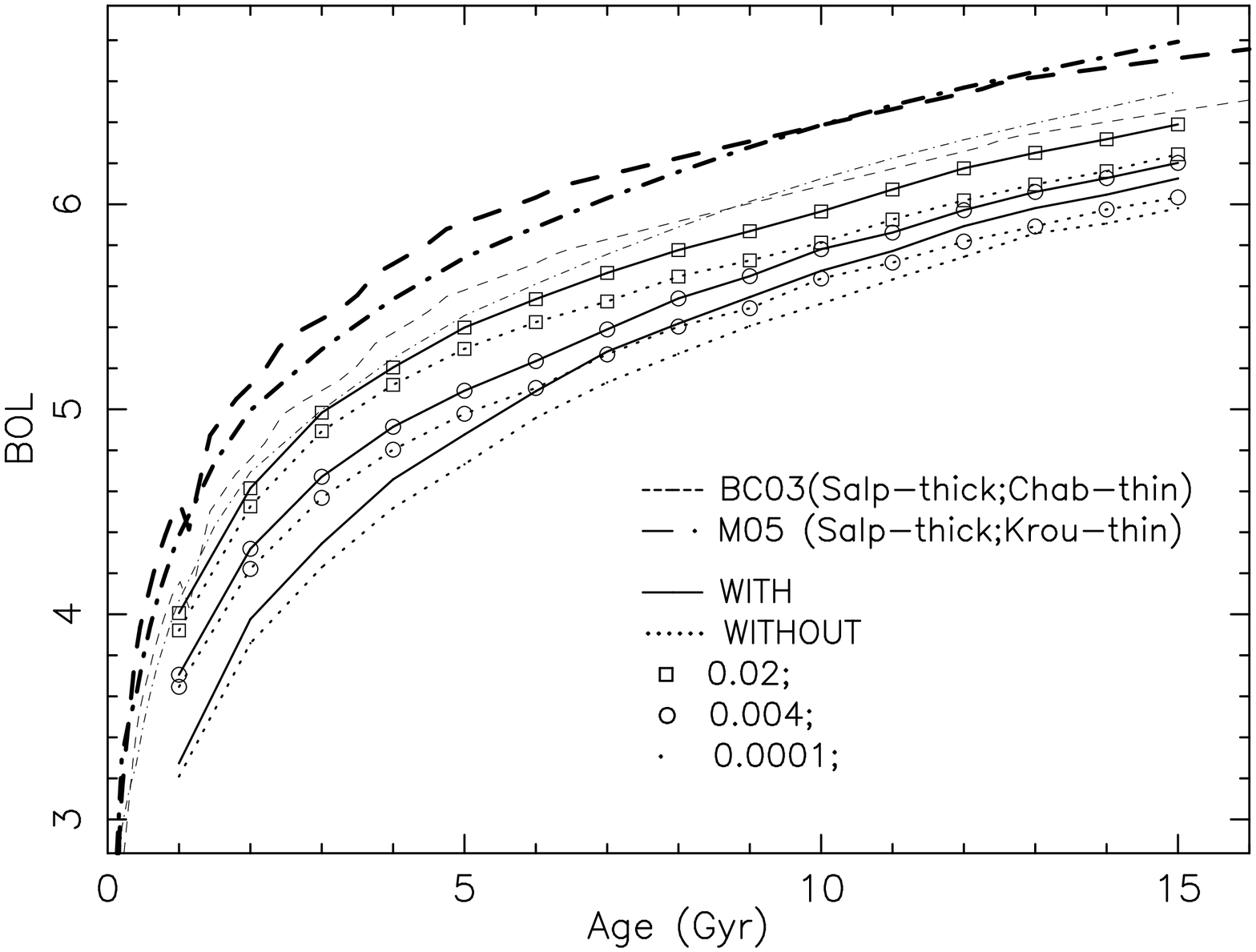}
\includegraphics[bb=79 37 581 702,height=3.0cm,width=2.6cm,clip,angle=-90]{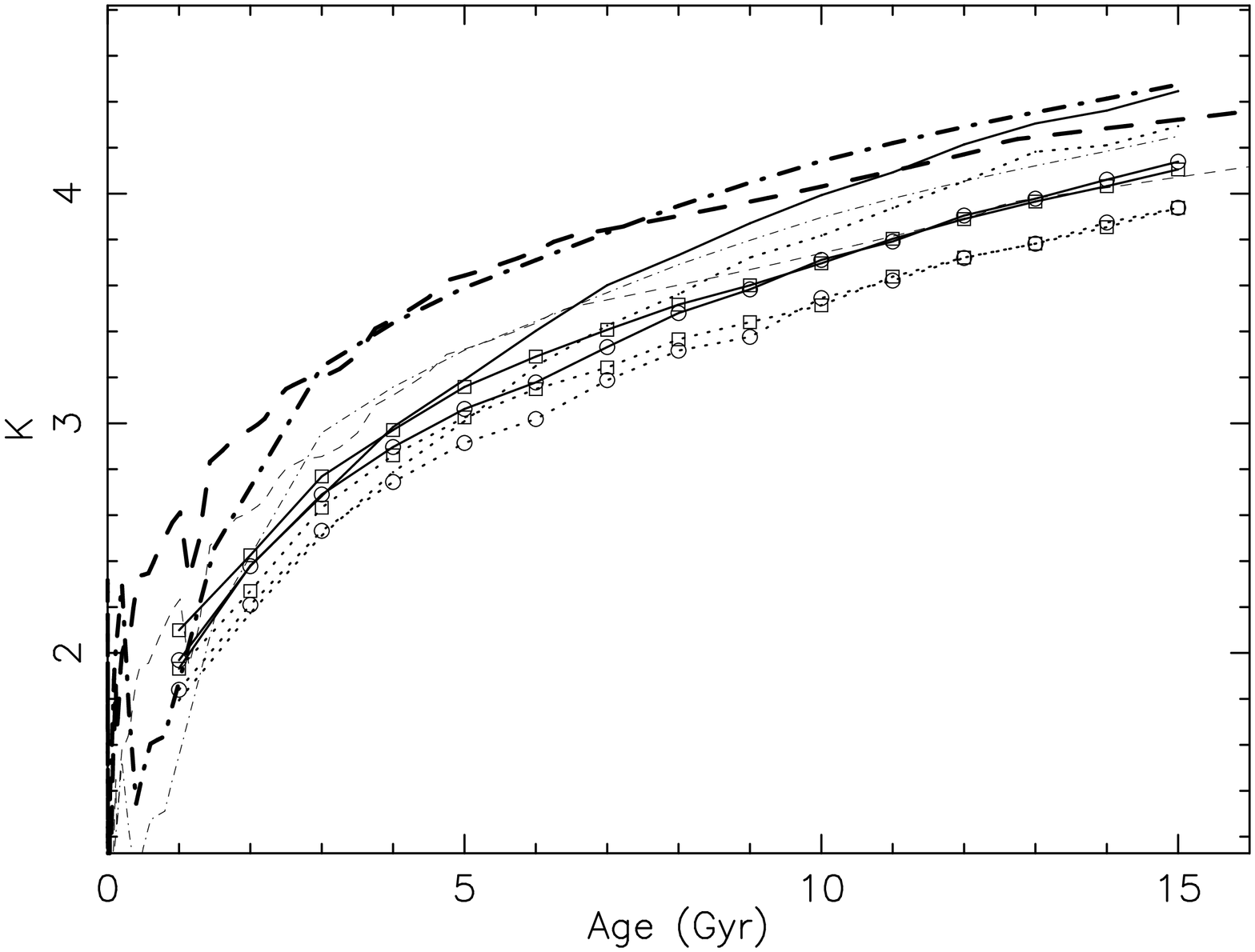}
}
\centering{
\includegraphics[bb=66 37 581 702,height=3.0cm,width=2.6cm,clip,angle=-90]{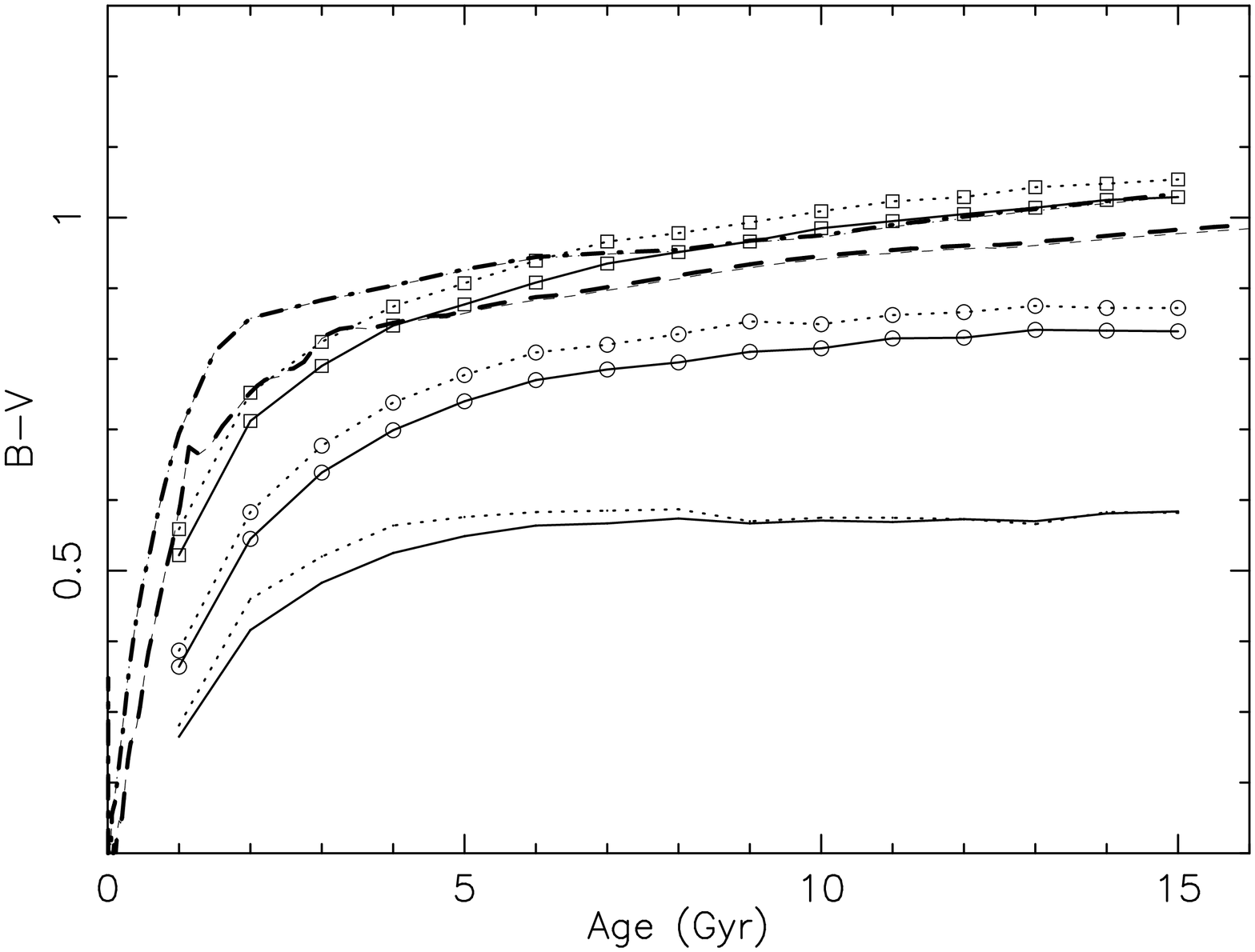}
\includegraphics[bb=66 37 581 702,height=3.0cm,width=2.6cm,clip,angle=-90]{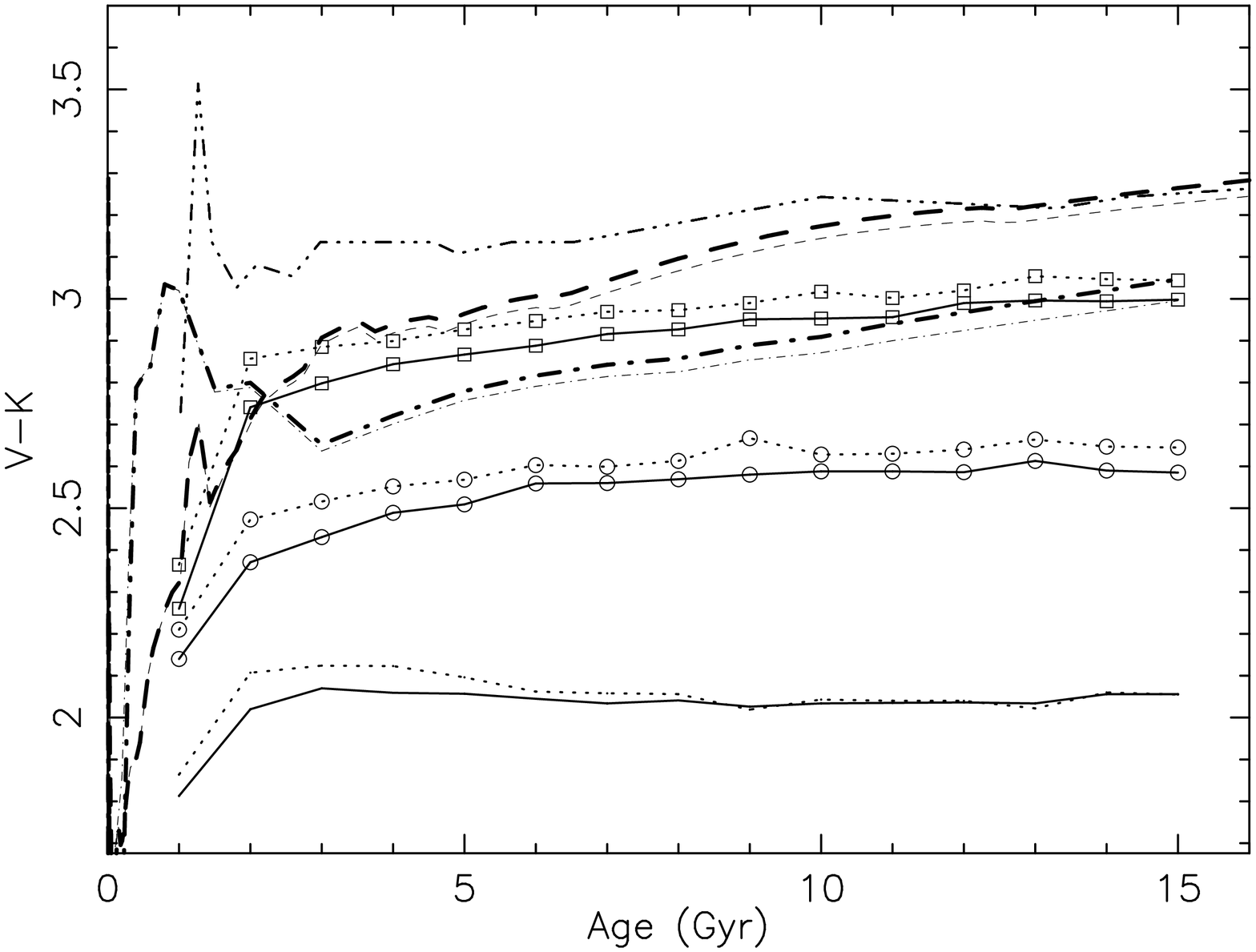}
}
\caption{~The bolometric magnitude ($M_{\rm BOL}$), $K$
magnitude, $B-V$ and $V-K$ colours for BSPs {\it WITH} and {\it
WITHOUT} BIs at $Z=0.02$, 0.04 and 0.0001. Also shown are the
results of BC03-S, BC03-C, M05-S and M05-K at solar metallicity.
In $V-K$ diagram the recent results of B07 are shown (dash-dot-dot-dot).}
\label{fig01}
\end{figure}

\begin{figure}
\centering{
\includegraphics[bb=135 87 567 626,height=3.0cm,width=2.6cm,clip,angle=-90]{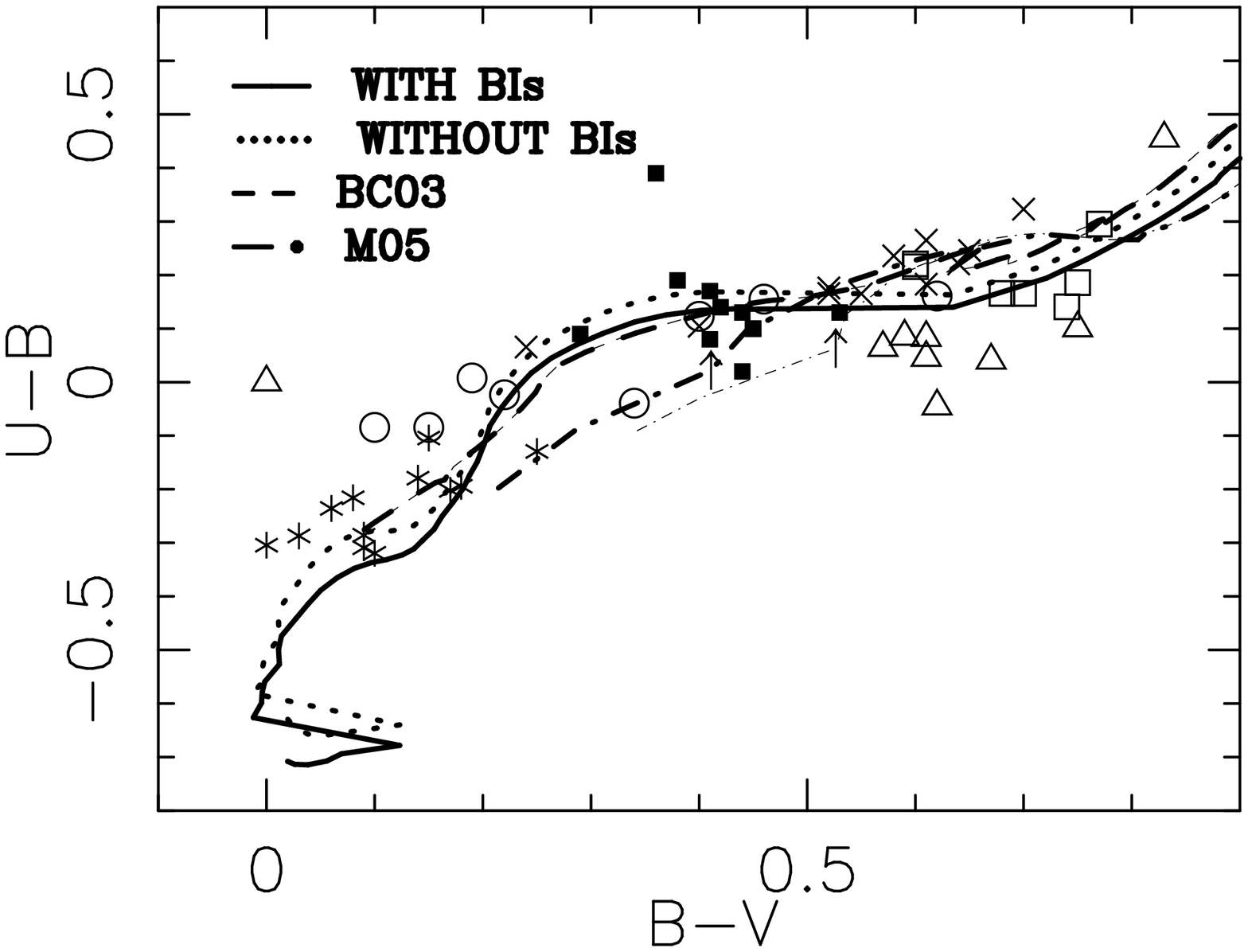}
\includegraphics[bb=135 87 567 626,height=3.0cm,width=2.6cm,clip,angle=-90]{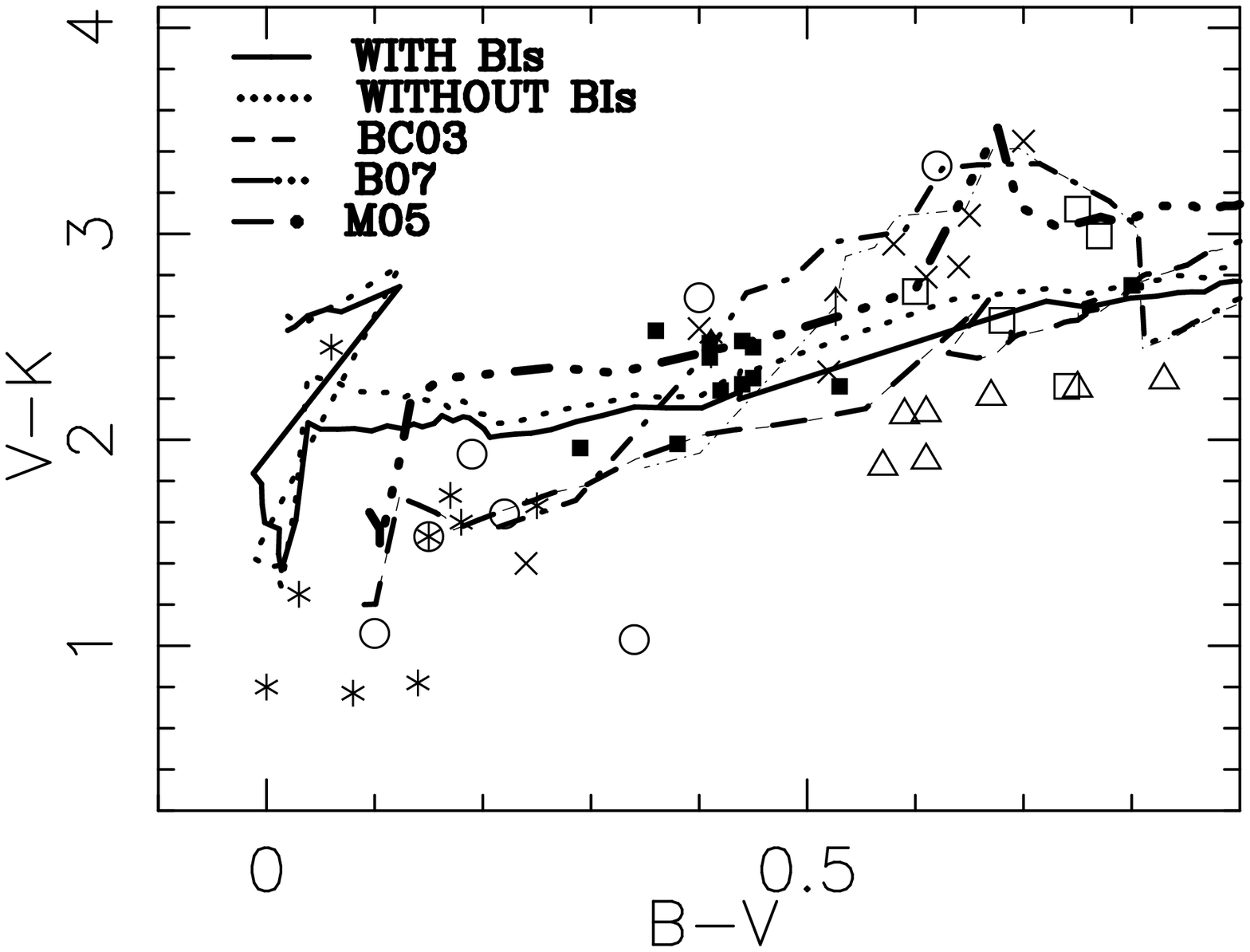}
}
\caption{~$B-V$ versus $U-B$, $B-V$ versus $V-K$ colours of star
clusters. The different symbols represent  Magellanic Clouds GCs
with the SWB type in the range 3-7. Solid rectangles show the
young star clusters in the merger remnant galaxy NGC 7252. The
full and dotted lines show the evolution of BSPs {\it WITH} and
{\it WITHOUT} BIs at $Z=0.01$, the ages of BSPs are
greater than a few Myr. The dashed lines are the BC03-S
(thick) and BC03-C (thin) models at $Z=0.008$, the dot-dash
lines are the M05-S (thick) and M05-K (thin) models at $Z=0.01$
with age $\rm{log} \tau >$ 8 yr, respectively. In $V-K$ diagram
the values of B07 at solar metallicity are also shown (dash-dot-dot-dot).}
\label{fig02}
\end{figure}

\begin{figure}
\centering{
\includegraphics[bb=118 95 563 625,height=3.0cm,width=2.5cm,clip,angle=-90]{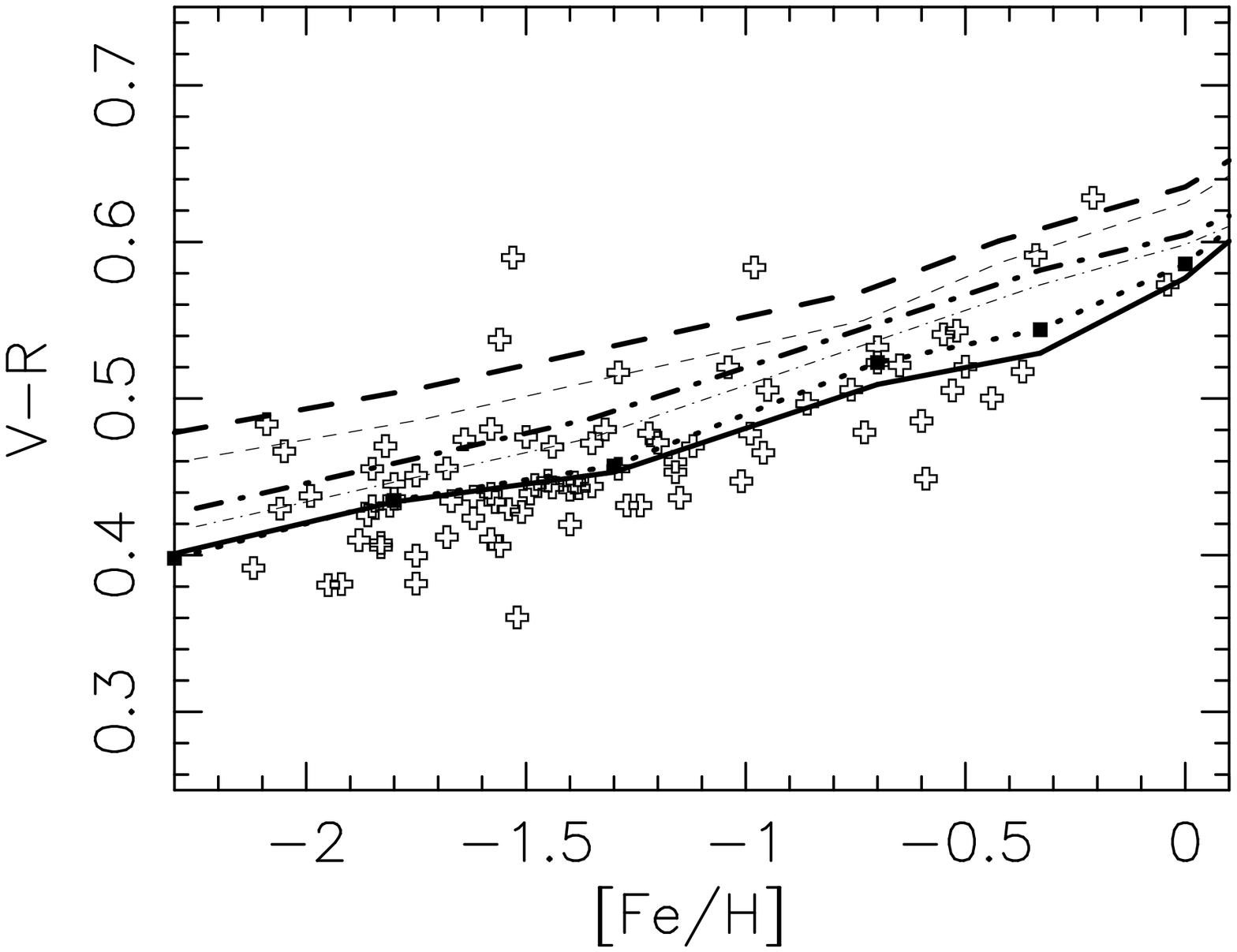}
\includegraphics[bb=118 95 563 625,height=3.0cm,width=2.5cm,clip,angle=-90]{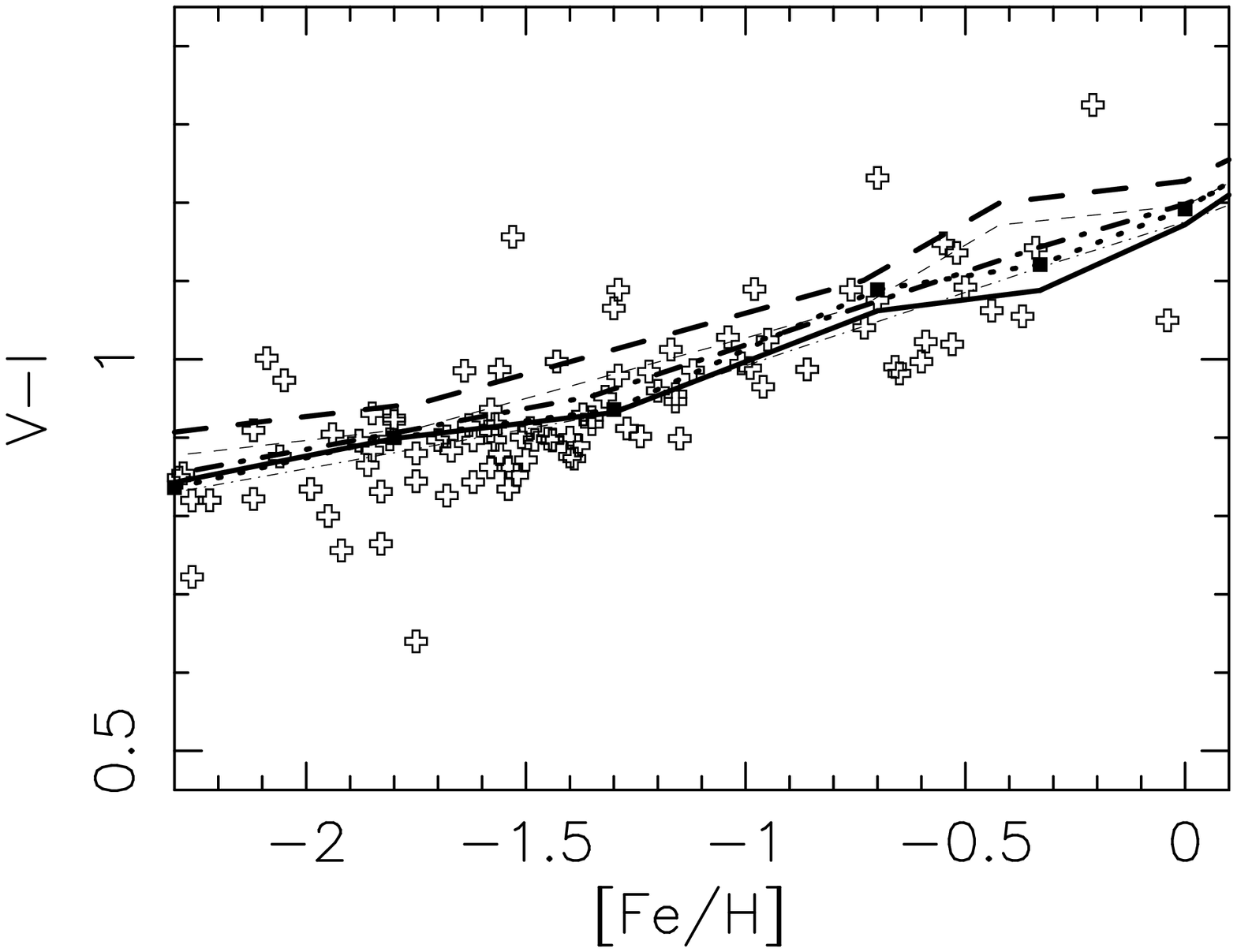}
\includegraphics[bb=118 95 563 625,height=3.0cm,width=2.5cm,clip,angle=-90]{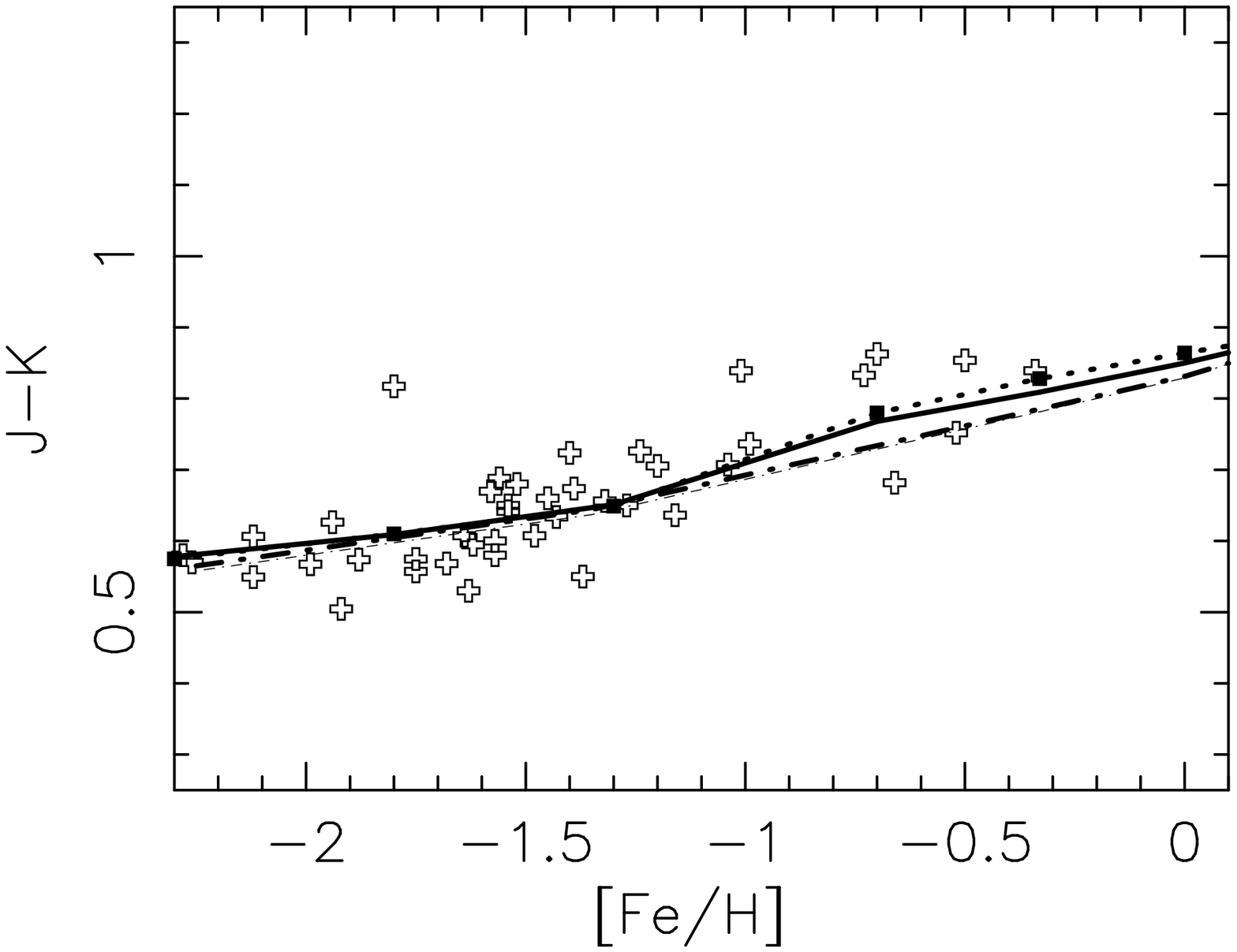}
}
\caption{~Comparison of colours with data of Milky Way GCs. The
lines represent the same models as in Fig. 2. All models have the
same age $\tau =$ 13\,Gyr}
\label{fig03}
\end{figure}

{\bf Results} We present the infrared integrated magnitudes and
colours for BSPs. The ages of BSPs are in the range 1-15\,Gyr, the
metallicities are in the range $0.0001-0.03$.

In Fig. 1 we present the bolometric magnitude $M_{\rm BOL}$, $K$
magnitude, $B-V$ and $V-K$ colours at $Z=0.02, 0.004$ and 0.0001
for BSPs {\it WITH} and {\it WITHOUT} BIs, the results of Bruzual
\& Charlot (2003, hereafter BC03) using Salpeter (1955, hereafter
BC03-S) and Chabrier (2003, hereafter BC03-C) IMFs, the results of
Maraston (2005, hereafter M05) using Salpeter (1955, hereafter
M05-S) and Kroupa (2001, hereafter M05-K) IMFs, and the recent
$V-K$ of Bruzual (2007, hereafter B07) at solar metallicity. By
comparison we see that {\bf (i)} the magnitudes and colours of
BSPs {\it WITH} BIs are greater (fainter) and smaller (bluer) than
those {\it WITHOUT} BIs, respectively. 
{\bf (ii)} The magnitudes of BC03-S, BC03-C, M05-S
and M05-K are greater than ours. {\bf (iii)} The shape of the
evolutionary curves of these colours is significantly different.
In Fig. 2 we compare the model colours with Magellanic Clouds
globular clusters (GCs) with the type of Searle, Wilkinson \&
Bagnuolo(1980, hereafter SWB) in the range of 3-7 and the young
star clusters in the merger remnant galaxy NGC 7252 in
$(B-V)$.vs.$(U-B)$ and $(B-V)$.vs.$(V-K)$ diagrams. It shows that
the BC03 and our models agree with the observations in
$(B-V)$.vs.$(U-B)$ diagram; while in $(B-V)$.vs.$(V-K)$ diagram
larger discrepancies exist among models.
In Fig. 3 we compare the model broad colours with Milky Way GCs in
colour-metallicity diagrams. It shows that our models match the
observations better than M05 and BC03 in ($V-R$)-[Fe/H] and
($V-I$)-[Fe/H] diagrams.

{\bf Acknowledgments} This work was funded by the Chinese Natural
Science Foundation (Grant Nos 10773026, 10673029, 10433030 \& 10521001)
and by Yunnan Natural Science Foundation (Grant Nos 2005A0035Q \&
2007A113M).

\end{document}